\documentclass[aps,pra,twocolumn,showpacs]{revtex4}%
\usepackage{amsmath}
\usepackage{amsfonts}
\usepackage{amssymb}
\usepackage{graphicx}%
\begin{document}
\title{Control of multiatom entanglement in a cavity}
\author{Aikaterini Mandilara and Vladimir M. Akulin}
\affiliation{Laboratoire Aime Cotton, CNRS, Campus d'Orsay, 91405, Orsay, France}
\author{Michal Kolar and Gershon Kurizki}
\affiliation{Weizmann Institute of Science, 76100 Rehovot, Israel}


\begin{abstract}
We propose a general formalism for analytical description of multiatomic
ensembles interacting with a single mode quantized cavity field under the
assumption that most atoms remain un-excited on average. By combining the
obtained formalism with the nilpotent technique for the description of
multipartite entanglement we are able to overview in a unified fashion
different probabilistic control scenarios of entanglement among atoms or
examine atomic ensembles. We then apply the proposed control schemes to the
creation of multiatom states useful for quantum information.

\end{abstract}
%\date[Date text ]{\today }
\maketitle

\section{Introduction}

Engineering of the entanglement properties of multipartite states has recently
become the subject of extensive research due to its relevance to quantum
information processing and computation. A number of entanglement control
schemes based on neutral atoms in a quantized cavity field \cite{Kimble,
Rempe, Raimond, determin} or in optical lattices \cite{optical} as well as
trapped-ion setups \cite{trapped} have been proposed, and implemented
experimentally. However, a universal theory of multipartite entanglement,
which could guide the diverse experimental and theoretical efforts, is still
missing. Instead, most of these proposals are focused on the generation of
specific entangled states\ rather than general states with an arbitrary chosen entanglement.

Here, we seek a general framework for multipartite entangled state engineering
of neutral atoms in a quantized cavity by making use of the recently
introduced formalism for the exhaustive description multipartite entanglement
based on the nilpotent polynomial technique \cite{nilpotent}.

The nilpotent formalism relies on raising operators acting on a reference
product state--for two-level systems the variables of these polynomials are
the nilpotent operators $\widehat{\sigma}_{i}^{+}$. Every quantum state can be
represented as a polynomial of operators $F(\{\widehat{\sigma}_{i}^{+}\})$
acting on a reference vector which we can choose to be the \textquotedblleft
atomic vacuum\textquotedblright\ $\left\vert \emph{O}\right\rangle =$ $%
%TCIMACRO{\dprod \limits_{n}}%
%BeginExpansion
{\displaystyle\prod\limits_{n}}
%EndExpansion
\left\vert 0\right\rangle _{n}$. The main object of interest is the logarithm
of $F(\{\widehat{\sigma}_{i}^{+}\})$, that we call \textquotedblleft
nilpotential\textquotedblright\ $f=\ln F(\{\widehat{\sigma}_{i}^{+}\})$. It
provides us with a simple criterion of entanglement for any binary partition,
$A$ and $B$, of the quantum system: the subsystems $A$ and $B$ are
disentangled iff $f_{A\cup B}=f_{A}+f_{B}$. By the term \textquotedblleft
tanglemeter\textquotedblright\ $f_{c}$ we denote the nilpotential of the
\textquotedblleft canonic state\textquotedblright\ which is the closest to the
atomic vacuum state among the variety of the states subject to all possible
local transformations. This polynomial contains exhaustive information about
the entanglement.

In order to implement the nilpotent polynomial technique to describe
controlled entanglement among \ $N$ neutral two-level atoms interacting with a
quantized single-mode electromagnetic field in a high-Q cavity, we further
assume that the number of excited atoms remains low on average, well below
$N$, during the interaction. In other words, we consider an entangled state of
a weakly excited multiatom ensemble that is coupled to the quantized field.
This assumption allows us to obtain an exact analytic description, which can
be directly interpreted in terms of the nilpotent formalism for multipartite
entanglement, and propose several techniques for rather general multiatom
entanglement engineering.

Unitary control over the system is exerted by squeezing, displacing, or
Kerr-nonlinear self-modulation of the cavity field, as well as manipulating
the coupling between the atoms and the field moving the atoms in the cavity.
This unitary control, combined with nonunitary measurements of the number of
cavity photons, that is projection of the\ cavity-field state onto a photon
number state, yields a vast variety of possible outcomes. We detail such a
probabilistic approach to the construction of the Dicke state for $N$
two-level atoms with $M$ excitations \cite{Kimble,Dicke}, the \textit{W} \cite{W}
\ \ and GHZ \cite{GHZ} states. We also present a procedure for
entangling two atomic ensembles. Decoherence is ignored throughout the
analysis. This is justified by the fact that in our analysis high-Q cavities
are required and also that the atoms remain low-excited on average.

The structure of the paper is the following. In Sec.\ref{separation} we derive
the analytic description of weakly excited atomic ensembles. In Sec.
\ref{methods} we turn to the entanglement control schemes, which we further
specify upon discussing their particular applications in Sec. \ref{secApp}. We
conclude by discussing the results obtained. Details of the calculations are
included in Appendices A and B.

\section{Analytic description of two-levels atoms in a
cavity\label{separation}}

In this section we derive an analytic expression for the time-dependent
entangled state of $N$ atoms and the cavity photons,with the help of the
functional integration technique, assuming that the number of atoms in the
excited state remains small as compared to $N$.

The Hamiltonian of a single-mode cavity field coupled to $N$ two-level atoms
via laser-induced Raman interaction reads, in the interaction representation,
\begin{equation}
\widehat{H}=\omega_{0}\widehat{a}^{\dagger}\widehat{a}+\sum_{n}\left[
\frac{\omega_{n}}{2}\widehat{\sigma}_{n}^{z}+\mathcal{E}(t)C_{n}\left(
t\right)  \left(  \widehat{\sigma}_{n}^{+}+\widehat{\sigma}_{n}^{-}\right)
\left(  \widehat{a}^{\dagger}+\widehat{a}\right)  \right]  . \label{EQ1}%
\end{equation}
Here $\omega_{0}$ and $\omega_{n}$ are the resonant frequencies of the cavity
mode and the $n$-th atom, respectively, $C_{n}(t)$ is the controlled Raman
coupling of the $n$-th atom with the cavity field induced by an external laser
field $\mathcal{E}(t)$, $\widehat{\sigma}_{n}^{z,+,-}$ are the Pauli matrices,
and $\widehat{a}^{\dagger}$, $\widehat{a}$ are the field-mode creation and
annihilation operators, respectively.

On separating the atomic and cavity field variables with the help of
a\ functional integral over the complex functional variable $z(t)$, we express
the evolution operator as
\begin{equation}
\widehat{U}=\int\frac{\mathcal{D}z(t)\mathcal{D}z^{\ast}(t)}{A}\exp\left[
-\mathrm{i}\int\left\vert z(t)\right\vert ^{2}\mathrm{d}t\right]  \widehat
{U}_{0}(t)\prod\limits_{n=1}^{N}\widehat{U}_{n}(t), \label{EQ2}%
\end{equation}
where the evolution operators $\widehat{U}_{n,o}(t)$ in the functional
integral satisfy the dynamic equations%
\begin{align}
\mathrm{i}\frac{\partial}{\partial t}\widehat{U}_{0}(t)  &  =\left[
\omega_{0}\widehat{a}\dagger\widehat{a}-\mathcal{E}(t)z^{\ast}(t)\left(
\widehat{a}\dagger+\widehat{a}\right)  \right]  \widehat{U}_{0}(t)
\label{EQ3}\\
\mathrm{i}\frac{\partial}{\partial t}\widehat{U}_{n}(t)  &  =\left[
\frac{\omega_{n}}{2}\widehat{\sigma}_{n}^{z}+z(t)C_{n}(t)\left(
\widehat{\sigma}_{n}^{+}+\widehat{\sigma}_{n}^{-}\right)  \right]  \widehat
{U}_{n}(t), \label{EQ4}%
\end{align}
and
\begin{equation}
A=\int\mathcal{D}z(t)\mathcal{D}z^{\ast}(t)\exp\left[  -\mathrm{i}%
\int\left\vert z(t)\right\vert ^{2}\mathrm{d}t\right]  \label{EQ5}%
\end{equation}
is a normalization factor.

For the field and the atoms initially in the ground states ($\left\vert
0\right\rangle _{p}$ and $\left\vert 0\right\rangle _{n}$, respectively), the
Schr\"{o}dinger equation gives
%TCIMACRO{\TeXButton{TeX field}{\begin{widetext}}}%
%BeginExpansion
\begin{widetext}%
%EndExpansion%
\begin{equation}
\left\vert \psi_{0}\left(  t\right)  \right\rangle =\exp\left[  \mathrm{i}%
\widehat{a}\dagger\int\limits_{0}^{t}\mathrm{e}^{\mathrm{i}\omega_{0}(\tau
-t)}\mathcal{E}(\tau)z^{\ast}(\tau)\mathrm{d}\tau-\int\limits_{0}^{t}%
\int\limits_{\tau}^{t}\mathrm{e}^{\mathrm{i}\omega_{0}(\tau-\theta
)}\mathcal{E}(\tau)\mathcal{E}(\theta)z^{\ast}(\tau)z^{\ast}(\theta
)\mathrm{d}\tau\mathrm{d}\theta\right]  \left\vert 0\right\rangle _{p}~~,
\label{EQ6}%
\end{equation}
while \ the second-order time-dependent perturbation theory yields, in the
interaction representation the quantum state
\begin{equation}
\left\vert \psi_{n}\left(  t\right)  \right\rangle \simeq\exp\left[
\mathrm{i}\frac{\omega_{n}}{2}t-\mathrm{i}\widehat{\sigma}_{n}^{+}%
\mathrm{e}^{-\mathrm{i}\omega_{n}t}\int\limits_{0}^{t}\mathrm{e}%
^{\mathrm{i}\omega_{n}\tau}C_{n}\left(  \tau\right)  z(\tau)\mathrm{d}%
\tau\right]  \left\vert 0\right\rangle _{n}. \label{EQ7}%
\end{equation}
The approximate expression (\ref{EQ7}) \ is valid as long as the mean number
of photons absorbed per atom remains small. With the help of Eqs.(\ref{EQ6}%
)-(\ref{EQ7}) \ and the functional integral Eq.(\ref{EQ2}), one finds the
quantum state of the compound atom-cavity system \
\begin{align}
\left\vert \Psi\right\rangle  &  =\int\frac{\mathcal{D}z(t)\mathcal{D}z^{\ast
}(t)}{A}\exp\left[  -\mathrm{i}\int\limits_{0}^{t}\left\vert z(\tau
)\right\vert ^{2}\mathrm{d}\tau+\mathrm{i}\sum\limits_{n=1}^{N}\frac
{\omega_{n}}{2}t\right]  \exp\left[  -\int\limits_{0}^{t}\int\limits_{\tau
}^{t}\mathrm{e}^{\mathrm{i}\omega_{0}(\tau-\theta)}\mathcal{E}(\tau
)\mathcal{E}(\theta)z^{\ast}(\tau)z^{\ast}(\theta)\mathrm{d}\tau
\mathrm{d}\theta\right] \nonumber\\
&  \exp\left[  -\mathrm{i}\sum\limits_{n=1}^{N}\widehat{\sigma}_{n}^{+}%
\int\limits_{0}^{t}\mathrm{e}^{\mathrm{i}\omega_{n}\left(  \tau-t\right)
}C_{n}\left(  \tau\right)  z(\tau)\mathrm{d}\tau\right]  \exp\left[
\mathrm{i}\widehat{a}\dagger\int\limits_{0}^{t}\mathrm{e}^{\mathrm{i}%
\omega_{0}(\tau-t)}\mathcal{E}(\tau)z^{\ast}(\tau)\mathrm{d}\tau\right]
\left\vert \mathcal{O}\right\rangle , \label{EQ8}%
\end{align}%
%TCIMACRO{\TeXButton{TeX field}{\end{widetext}} }%
%BeginExpansion
\end{widetext}
%EndExpansion
where by $\left\vert \mathcal{O}\right\rangle $ we denote the
\textquotedblleft vacuum state\textquotedblright\ \ $\left\vert 0\right\rangle
_{p}%
%TCIMACRO{\dprod \limits_{n}}%
%BeginExpansion
{\displaystyle\prod\limits_{n}}
%EndExpansion
\left\vert 0\right\rangle _{n}$ henceforth. Details of the functional
integration are presented in Appendix A, and the final result Eq.(\ref{EQ17})
is given by the following explicit expression%

\begin{align}
\left\vert \Psi\left(  t\right)  \right\rangle  &  =\frac{1}{A(t)}\exp\left[
\widehat{a}\dagger\sum\limits_{n=1}^{N}\widehat{\sigma}_{n}^{+}I_{n}\right]
\nonumber\\
&  \exp\left[  \sum\limits_{n,m=1}^{N}\widehat{\sigma}_{n}^{+}\widehat{\sigma
}_{m}^{+}I_{n,m}\right]  \left\vert \mathcal{O}\right\rangle , \label{EQ18a}%
\end{align}
with the coefficients $I$ given as
\begin{align}
I_{n}  &  =\mathrm{i}\int\limits_{0}^{t}\mathrm{e}^{\mathrm{i}\left(
\omega_{0}+\omega_{n}\right)  (\tau-t)}C_{n}\left(  \tau\right)
\mathcal{E}(\tau)\mathrm{d}\tau,\label{18a}\\
I_{n,m}  &  =\int\limits_{0}^{t}\int\limits_{0}^{\tau}C_{n}\left(
\tau\right)  \mathcal{E}(\tau)C_{m}\left(  \theta\right)  \mathcal{E}%
(\theta)\nonumber\\
&  \mathrm{e}^{-\mathrm{i}\left(  \tau(\omega_{0}-\omega_{m}\right)
+t(\omega_{n}+\omega_{m)}-\theta(\omega_{n}+\omega_{0}))}\mathrm{d}%
\theta\mathrm{d}\tau, \label{18b}%
\end{align}
and $A(t)$ being the time-dependent normalization.

Eq. (\ref{EQ18a}) gives the general form of an entangled quantum state for a
quantized cavity field mode coupled to an ensemble of $N$ low-excited atoms.
\ The presence of the oscillating terms in the integrals of Eqs.(\ref{18a}%
)-(\ref{18b}) \ guarantee that the parameters $I_{n}$ and $I_{n,m}$ are
typically small numbers and therefore the probability of excitation per atom
remains low. However, when $\omega_{0}\simeq-\omega_{n}$, in the cavity-atom
resonant regime, the coefficients $I_{n}$ can be significantly larger as
compared to the negligible $I_{n,m}$ parameters. \ As we show in
Sec.\ref{secApp} in detail\ the probability for each atom to be in the excited
state in this case indeed remain small.

As the next step we establish the connection between the parameters in
Eqs.(\ref{18a})-(\ref{18b}), the \ measured number of the cavity photons, and
the nilpotential's coefficients characterizing the entanglement in the atomic ensemble.

\section{Methods of entanglement control\label{methods}}

We are now in a position to relate the quantum state Eq.(\ref{EQ18a}) to the
nilpotential formulation introduced in Ref.\cite{nilpotent}. Cavity photons
\ can be incorporated into the description, with the raising operator
$\widehat{a}\dagger$ being the corresponding `nilpotent' variable. In
particular, for the state in Eq.(\ref{EQ18a}), the tanglemeter $f_{c}$,
defined as $\left\vert \Psi\right\rangle =$ $\frac{1}{A(t)}\mathrm{e}^{f_{c}%
}\left\vert \mathcal{O}\right\rangle $, has the form%

\begin{equation}
f_{c}=\widehat{a}\dagger\sum\limits_{n=1}^{N}\widehat{\sigma}_{n}^{+}%
I_{n}+\sum\limits_{n,m=1}^{N}\widehat{\sigma}_{n}^{+}\widehat{\sigma}_{m}%
^{+}I_{n,m}\ . \label{ent}%
\end{equation}
According to the entanglement criterion, the first term describes the
entanglement among the atoms and the photons whereby the atoms collectively
participate in this entangled photons-atoms state via the operator
$\sum\limits_{n=1}^{N}\widehat{\sigma}_{n}^{+}I_{n}$. This term corresponds to
the lowest order of entanglement between the collective state of \ the atoms
and the cavity photons. The second term in Eq.(\ref{ent})\ is concerned
exclusively with the entanglement among the atoms. In this work we are mostly
interested in understanding how the atoms get entangled after the degrees of
freedom of photons are traced out by a projective measurement on the
\textquotedblleft engineered\textquotedblright\ cavity field state. After the
measurement of the cavity photon number, the tanglemeter of the multiatomic
system undergoes a transformation to the generic form%
\begin{equation}
f_{c}=\sum\limits_{n,m=1}^{N}\beta_{n,m}^{\left(  2\right)  }\widehat{\sigma
}_{n}^{+}\widehat{\sigma}_{m}^{+}+\ldots+\sum\limits_{n,m,\ldots,l=1}^{N}%
\beta_{n,m\ldots,l}^{\left(  N\right)  }\widehat{\sigma}_{n}^{+}%
\widehat{\sigma}_{m}^{+}\ldots\widehat{\sigma}_{l}^{+} \label{p20}%
\end{equation}
where higher order terms imply the presence of higher order entanglement. For
instance, a GHZ state of \textit{N }two-levels atoms requires a
non-zero term proportional to $\beta^{\left(  N\right)  }$ while bipartite
entanglement requires only the second-order terms to be nonzero.

In principle, all the $\beta$ coefficients in Eq.(\ref{p20}) are different and
hard to control. We shall therefore consider a simpler problem of entanglement
between two multiatomic ensembles, $A$ and $B$, each containing $N$ atoms,
equivalent from the viewpoint of entanglement.

To describe the entanglement between the two ensembles, we define for each
ensemble a collective nilpotent variable: $\widehat{\sigma}_{A}^{+}%
=\sum\limits_{n=1}^{N}\widehat{\sigma}_{n}^{\left(  A\right)  +}$ and
$\widehat{\sigma}_{B}^{+}=\sum\limits_{n=1}^{N}\widehat{\sigma}_{n}^{\left(
B\right)  +}$ respectively. These variables are not nilpotent in the limit
$N\rightarrow\infty$ since they only vanish in the power $N+1$. Still,\ the
collective operator \ $\widehat{\sigma}_{A}^{+}$ together with the operators
$\widehat{\sigma}_{A}^{-}$ and $\widehat{\sigma}_{A}^{z}$ form the
$\mathbf{su}(2)$ algebra that is a subalgebra of the full algebra
$\mathbf{su}(2^{N})$ of the $N$-atom ensemble. This situation belongs to the
case of generalized entanglement \cite{Lorenza}, which admits powers higher
than two of the creation operators. The tanglemeter for two ensembles has then
the general form%
\begin{equation}
f_{c}=\overset{N}{\underset{k,l=0}{\sum}}\beta_{k,l}\left(  \widehat{\sigma
}_{A}^{+}\right)  ^{k}\left(  \widehat{\sigma}_{B}^{+}\right)  ^{l}\ .
\label{p21}%
\end{equation}
The presence of higher-order cross terms is associated with higher-order
entanglement between the two ensembles, while nonlinear terms that are
dependent only on one nilpotent variable refer to the entanglement among atoms
within each ensemble.

We proceed by listing different methods for entanglement control, namely, the
ways of manipulating the tanglemeter coefficients $\beta$ of Eq.(\ref{p20})
for a multiatomic ensemble, or that of Eq.(\ref{p21}) for two such ensembles.

\subsection{Control of the coefficients in Eq.(\ref{ent}).}

The $\beta$ coefficients in the tanglemeter of Eq.(\ref{p20}) directly depend
on the initial combined atoms-photons state in Eq.(\ref{EQ18a}). Therefore,
prior to the field manipulations and measurement, one can affect the final
state by controlling the $N(N+1)/2$ coefficients $I_{n}$ and $I_{n,m}$ in the
nilpotential Eq.(\ref{ent}). According to Eqs.(\ref{18a})-(\ref{18b}), these
coefficients depend on time-integrals over the prescribed time-dependent laser
field $\mathcal{E}(t)$ \ and on the coupling parameters $C_{n}(t)$. The latter
parameters are determined by the cavity geometry, and the individual positions
of the atoms inside the cavity. The prospects for realizing this crucial requirement  by emerging techniques \cite{Merschede} are discussed in Sec. \ref{V}.

One has complete control over the coefficients $I_{n,m}$, when the number of
adjustable parameters of the laser field \ and the couplings exceeds the total
number $N(N+1)/2$ of these coefficients. For the simplest control setting,
this implies that a constant field $\mathcal{E}$ is switched on and off
$N(N+1)/2$ times, such that each time when the field is on, only one of
$N(N+1)/2$ possible pairs of the atoms is in the cavity. Given the
coefficients $I_{n}$ and $I_{n,m}$, one finds the time intervals when the
field is on by solving standard linear algebraic equations. This approach
holds for arbitrary shapes of $\mathcal{E}(t)$ and \ $C(t)$ cast into a
superposition of $N(N+1)/2$ linearly-independent functions of time, as long as
the corresponding linear problem is not singular.

Finally, one can control the relative strength of the linear $I_{n}$ and
bilinear $I_{n,m}$ coefficients by adjusting the \ relative frequencies,
$\omega_{0}+\omega_{n}$ and $\omega_{0}-\omega_{n}$, \ in the oscillating
terms \ of the integrals in Eqs.(\ref{18a})-(\ref{18b}). For instance, in
Sec.\ref{secApp} control schemes are proposed in the cavity-atoms resonant
regime, $\omega_{0}\simeq-\omega_{n}$, where the bilinear terms $I_{n,m}$ in
Eq.(\ref{18b}) become negligible compared to the $I_{n}$ terms.

\subsection{Measuring the field in the cavity.}

One of the possibilities to control the atomic entanglement is by measuring
the number of photons in the cavity \cite{Kimble, Molmer, Plenio, Chou05}. The
probabilistic outcome after having measured $d$ photons yields the non
normalized state
\begin{equation}
\left\vert \Phi\right\rangle =\left(  \sum\limits_{n=1}^{N}\widehat{\sigma
}_{n}^{+}I_{n}\right)  ^{d}\exp\left[  \sum\limits_{n,m=1}^{N}\widehat{\sigma
}_{n}^{+}\widehat{\sigma}_{m}^{+}I_{n,m}\right]  \left\vert \emph{O}%
\right\rangle \ .\label{EQ21}%
\end{equation}

In the special case of the cavity vacuum, $d=0$, the nilpotential of the
resulting atomic state is \ already in the tanglemeter form $f_{c}%
=\sum\limits_{n,m=1}^{N}\widehat{\sigma}_{n}^{+}\widehat{\sigma}_{m}%
^{+}I_{n,m}$. The presence of only bilinear terms in the tanglemeter indicates
that the degree of entanglement is not high. Only for three \cite{Wootters, W}
and four two-level atoms \cite{Verstraete, nilpotent} can one can prove that
this state contain ingenious three-partite and four-partite entanglement,
respectively. In general, for constructing higher-entangled states, one needs
either to detect $d\geq1$ photons, or to perform certain manipulations\ of the
field prior to detecting $d=0$ photons. We dwell on the latter scenario in the
next paragraphs by considering realistic manipulations on optical cavities.
The techniques to be presented are hardly practical for superconducting
microwave cavities where the field of the cavity is inaccessible
\cite{Raimond}.

\subsection{Displacing the cavity field prior to the measurement.}

Let us first apply the field displacement operator $D=\mathrm{e}^{\frac{1}%
{2}\left\vert \lambda\right\vert ^{2}}\mathrm{e}^{-\lambda^{\ast}%
\widehat{\alpha}}$ $\mathrm{e}^{\lambda\widehat{\alpha}^{\dag}}$to the state
in Eq.(\ref{EQ18a}) that experimentally implies injecting a classical field
into the cavity \cite{Scully}, and then measure the cavity photon number. If
$d=0$ is detected, we keep the projected atomic state, otherwise we discard
it. This yields the non normalized state%
\begin{equation}
\ \left\vert \Phi\right\rangle =\exp\left[  -\lambda^{\ast}\sum\limits_{n=1}%
^{N}\widehat{\sigma}_{n}^{+}I_{n}\right]  \exp\left[  \sum\limits_{n,m=1}%
^{N}\widehat{\sigma}_{n}^{+}\widehat{\sigma}_{m}^{+}I_{n,m}\right]  \left\vert
\emph{O}\right\rangle \ .
\end{equation}
The local operator $\exp\left[  -\lambda^{\ast}\sum\limits_{n=1}^{N}%
\widehat{\sigma}_{n}^{+}I_{n}\right]  $ is nonunitary. In combination with the
local unitary $SU(2)$ transformations it allows one to perform all
transformations of the group $SL(2,%
%TCIMACRO{\U{2102} }%
%BeginExpansion
\mathbb{C}
%EndExpansion
)$. Therefore, by displacing the cavity field, performing local unitary
operations and measurements of the photon number, we can move the state
Eq.(\ref{EQ21}) \ with $d=0$, along its $SL$-orbit \cite{Verstraete} thus
changing the amount of $su$-entanglement (see Ref.\cite{nilpotent} for more
details). The operations belonging to the $SL(2,%
%TCIMACRO{\U{2102} }%
%BeginExpansion
\mathbb{C}
%EndExpansion
)$ group can be used for entanglement distillation \cite{Verstraete2}. For
three two-level atoms (qubits), for example, one can use displacement of the
field to construct with some probability the  GHZ state. For more than
three atoms the $SL$-orbit of the state Eq.(\ref{EQ21}) \ with $d=0$ does not
contain the GHZ state and therefore cannot be obtained by this
method\ -- the fidelity between the GHZ state and the closest state
one can construct by this method is decreasing rapidly with the number of atoms.

\subsection{Squeezing the cavity field prior to the measurement.}

Squeezing the cavity field state by applying the operator$\ \exp\left[
\left(  g\widehat{\alpha}^{2}-g^{\ast}\widehat{\alpha}^{\dagger2}\right)
t\right]  $ prior to the photon number measurement offers another possibility
to control the atomic entanglement. Experimentally, this amounts to operating
the gas-loaded cavity as a parametric amplifier. If we start, as earlier, with
the state Eq.(\ref{EQ18a})
\begin{equation}
\left\vert \Psi\right\rangle =\exp\left[  \widehat{a}\dagger\widehat
{O}+\widehat{G}\right]  \left\vert \mathcal{O}\right\rangle \
\end{equation}
with $\widehat{O}=$ $\sum\limits_{n=1}^{N}\widehat{\sigma}_{n}^{+}I_{n}$ and
$\widehat{G}=\sum\limits_{n,m=1}^{N}\widehat{\sigma}_{n}^{+}\widehat{\sigma
}_{m}^{+}I_{n,m}$ and detect zero photons in the cavity after the squeezing,
the reduced state of the multiatom ensemble reads
\begin{align}
\left\vert \Phi\right\rangle  &  =\left\langle 0\right\vert _{p}%
\mathrm{e}^{\left(  g\widehat{\alpha}^{2}-g^{\ast}\widehat{\alpha}^{\dagger
2}\right)  t}\mathrm{e}^{\widehat{a}\dagger\widehat{O}+\widehat{G}}\left\vert
0\right\rangle _{p}\left\vert \emph{O}\right\rangle \nonumber\\
&  =\left\langle 0\right\vert _{p}\mathrm{e}^{-\widehat{a}\dagger\widehat{O}%
}\mathrm{e}^{\left(  g\widehat{\alpha}^{2}-g^{\ast}\widehat{\alpha}^{\dagger
2}\right)  t}\mathrm{e}^{\widehat{a}\dagger\widehat{O}}\mathrm{e}^{\widehat
{G}}\left\vert 0\right\rangle _{p}\left\vert \emph{O}\right\rangle \nonumber\\
&  =\left\langle 0\right\vert _{p}\mathrm{e}^{g\left(  \widehat{\alpha
}+\widehat{O}\right)  ^{2}t-g^{\ast}\widehat{\alpha}^{\dagger2}t+\widehat{G}%
}\left\vert 0\right\rangle _{p}\left\vert \emph{O}\right\rangle .\label{40}%
\end{align}
For a real squeezing parameter $g$, the non normalized state Eq.(\ref{40}) can
be rewritten as $\left\vert \Phi\right\rangle =\exp(\zeta\widehat{O}%
^{2}+\widehat{G})\left\vert \emph{O}\right\rangle ,$ (see Appendix \ref{appB}
for details). The presence of the square of the operator $\widehat{O}$ in the
atoms' nilpotential \ implies that the atoms can be entangled with each other
even if \ zero photons are detected and the cavity-atoms frequencies\ are
tuned to the resonant regime where $\widehat{G}$ is negligible. In particular,
this can be done for two ensembles as shown in Sect.\ref{secApp}.

\subsection{Creation of highly entangled atomic states by field nonlinearity}

Apart from displacement and squeezing there exists another tool for the field
manipulation: a Kerr-nonlinear gas (with appreciable nonlinearity at the
cavity-mode frequency $\omega_{0}$) can be introduced in the cavity after the
atoms have passed through it. This type of coupling results in a nonlinear
dependence of the cavity energy on the number of the cavity photons. The
presence of the medium can also couple an external laser field at a frequency
$\omega_{L}$ with the cavity field, thus inducing a multiphoton cavity excitation.

To be more specific, let us consider a symmetric \textit{Kerr} medium with the
nonlinear polarization $\mathbf{P}=\xi\mathbf{E}^{3}$, where the electric
field $\mathbf{E}$ consists of the classical laser field of \textit{large}
amplitude $\mathcal{E}$ and the quantum cavity field $\sim\left(  \widehat
{a}\dagger+\widehat{a}\right)  $. \ In the rotating frame defined by the
unitary transformation $\widehat{U}_{0}=\exp\left[  -\mathrm{i}\omega
_{L}\widehat{a}\dagger\widehat{a}t\right]  $ , the interaction energy
$\mathbf{PE}$ integrated over the cavity volume $\mathcal{V}$ yields the
Hamiltonian
\begin{equation}
\widehat{H}_{c}\simeq\left(  \omega_{0}-\omega_{L}\right)  \widehat{a}%
\dagger\widehat{a}+\kappa\left[  \left(  \widehat{a}\dagger\widehat{a}\right)
^{2}+\left(  \widehat{a}\dagger+\widehat{a}\right)  \mathcal{E}^{3}\right]  ,
\label{27}%
\end{equation}
with $\kappa\sim\xi\mathcal{V}$. \ By adjusting the laser frequency
$\omega_{L}$ and amplitude $\mathcal{E}$, one finds a multiphoton resonance
with the cavity photons and creates superpositions of the cavity photon states
that differ by a large fixed number of photons. If one detects the cavity in
the vacuum state after the initial state Eq.(\ref{EQ18a}) has been subject to
such a transformation, the atomic ensemble turns out to be in a highly
entangled state. This can become a GHZ-state for certain values of the
parameters, as it will be shown in the next section (Sec. \ref{IVb}).

\section{Applications\label{secApp}}

Having presented the main ideas of the control schemes we illustrate the
suggested techniques for specific states, such as the\textit{ }$N$%
\textit{-}atom GHZ \ and the $N$\textit{-}atom Dicke state with $M$
excitations. The $W$ state of $N$\textit{-}atom is  a Dicke state with $M=1$.
The state vectors of these states read%
\begin{subequations}
\begin{align}
\left\vert GHZ\right\rangle  &  =\frac{\left\vert 00\ldots0\right\rangle
+\left\vert 11\ldots1\right\rangle }{\sqrt{2}};\nonumber\\
\quad\left\vert W\right\rangle  &  =\left\vert 1,N\right\rangle _{D}%
=\frac{\sum_{k}P_{k}\left\vert 10\ldots0\right\rangle }{\sqrt{N}};\nonumber\\
\left\vert M,N\right\rangle _{D} &  =\sqrt{\frac{M!\left(  N-M\right)  !}{N!}%
}\underset{k}{%
%TCIMACRO{\tsum }%
%BeginExpansion
{\textstyle\sum}
%EndExpansion
}P_{k}\underset{M}{|\underbrace{11\ldots1}0}\ldots00>,\label{36}%
\end{align}
\ respectively, where $\left\{  P_{k}\right\}  $ denotes the set of all
distinct permutations of atoms. The corresponding tanglemeters are of the
form
\begin{align}
f_{GHZ} &  =\sigma_{1}^{+}\sigma_{2}^{+}\ldots\sigma_{N}^{+};\quad
f_{W}=\underset{i=1,i\neq j}{\overset{N}{%
%TCIMACRO{\tsum }%
%BeginExpansion
{\textstyle\sum}
%EndExpansion
}}\sigma_{j}^{+}\sigma_{i}^{+};\label{tang}\\
f_{D} &  =\overset{M}{\underset{j=1}{%
%TCIMACRO{\tsum }%
%BeginExpansion
{\textstyle\sum}
%EndExpansion
}}\sigma_{j}^{+}\underset{i=M+1}{\overset{N}{%
%TCIMACRO{\tsum }%
%BeginExpansion
{\textstyle\sum}
%EndExpansion
}}\sigma_{i}^{+}+\underset{\underset{k\neq m}{k,m=1}}{\overset{M}{%
%TCIMACRO{\tsum }%
%BeginExpansion
{\textstyle\sum}
%EndExpansion
}}\sigma_{k}^{+}\sigma_{m}^{+}\underset{\underset{i\neq j}{i,j=M+1}}%
{\overset{N}{%
%TCIMACRO{\tsum }%
%BeginExpansion
{\textstyle\sum}
%EndExpansion
}}\sigma_{i}^{+}\sigma_{j}^{+}\nonumber\\
&  +\ldots+\underset{i=1}{\overset{M}{%
%TCIMACRO{\tprod }%
%BeginExpansion
{\textstyle\prod}
%EndExpansion
}}\sigma_{i}^{+}\underset{j,k,\ldots l=M+1}{\overset{N}{%
%TCIMACRO{\tsum }%
%BeginExpansion
{\textstyle\sum}
%EndExpansion
}}\underset{M}{\underbrace{\sigma_{j}^{+}\sigma_{k}^{+}\ldots\sigma_{l}^{+}}%
.}\nonumber
\end{align}
These states are often discussed in the context of quantum information
processing. Moreover, a prominent experimental realization of quantum
continuous variables is obtained, when the quantum states of an asymptotically
large ensemble are treated collectively, admitting operators with
asymptotically continuous spectra \cite{continuous}. Therefore one of our
examples is concerned with the engineering of entanglement between two atomic ensembles.

We note that all the constructions to be presented are probabilistic in nature
and the desired state is attained \ only when the cavity field is detected in
a specific Fock state. Yet, the success rate can be appreciable. Even though
the methods presented so far concern the general case where all the $I$
coefficients Eqs.(\ref{18a})-(\ref{18b}) are nonzero, we shall next consider
applications in the resonant cavity-atoms regime where the $I_{n,m}$ are
negligible, while the $I_{n}$ are not.

\subsection{Constructing a Dicke state with M excitations}

Consider $N$ identical atoms that are sent through the cavity of resonant
frequency $\omega_{0}\simeq-\omega_{n}$. If all the atoms are manipulated
equivalently, the combined wave function Eq.(\ref{EQ18a}) yields the state
\end{subequations}
\begin{equation}
\left\vert \Psi\left(  t\right)  \right\rangle =\frac{1}{A}\exp\left[
\widehat{a}\dagger c\sum\limits_{n=1}^{N}\widehat{\sigma}_{n}^{+}\right]
\left\vert \mathcal{O}\right\rangle \ , \label{30}%
\end{equation}
with identical field-atom coupling coefficients $I_{i}=c$, and vanishing
pairwise coefficients $I_{i,j}=0$. The normalization factor $A$ is\
\begin{equation}
A=\sum\limits_{i=1}^{N}\frac{i!N!}{(N-i)!}\ \left\vert c\right\vert ^{2i},
\label{31}%
\end{equation}
while the excitation probability for each atom $\ $is $1/N$. \ Therefore, for
large $N$, Eq.(\ref{30}) is consistent with our initial assumption of low
excitation per atom.

When a measurement of the cavity photon number is performed, the result $d=M$
is obtained with probability%
\begin{equation}
P(M,c)=\frac{\frac{M!N!}{(N-M)!}\ \left\vert c\right\vert ^{2M}}%
{\sum\limits_{i=1}^{N}\frac{i!N!}{(N-i)!}\ \left\vert c\right\vert ^{2i}},
\label{22}%
\end{equation}
leaving the atomic ensemble in the $\left\vert M,N\right\rangle _{D}$\textit{
}state. Experimentally one can adjust the coupling coefficient $I_{i}=c$ such
as the probability of attaining the desired Dicke state becomes maximum. In
Fig. 1 we show for the ensembles of $N=10$ and $N=19$ atoms, the probability
$P(M,c)$ \ Eq.(\ref{22}) for different values of the coupling coefficient
$\left\vert c\right\vert $ and for different excitations $M$. \ We\ note that
the Dicke states $\left\vert M,N\right\rangle _{D}$ and $\left\vert
N-M,N\right\rangle _{D}$ although they are equivalent up to local operation on
each atom, the maximum probability for the creation of each of them might be
considerably different. Therefore, one should choose to construct the one that
acquires highest probability $P$.

\begin{figure}[th]
{\centering{\includegraphics*[width=0.5\textwidth]{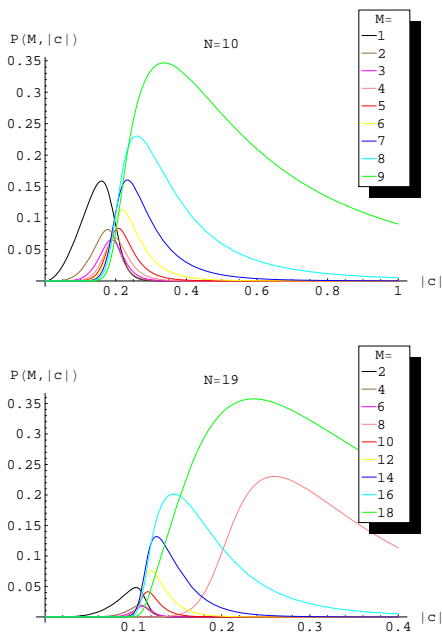}}} \vspace
{0.1cm}\caption{(Color online)  The probability $P(M,c)$ for obtaining the Dicke state
$\left\vert M,N\right\rangle _{D}$ as a function of the amplitude of the
coupling coefficient $\left\vert c\right\vert $. We consider two ensembles
consisting of $N=10$ and $N=19$ atoms, and a different number of excitations
$M$. }%
\label{Fig.1}%
\end{figure}

\subsection{Construction of the GHZ state of an atomic ensemble}

\label{IVb}

We begin with the observation that the state Eq.(\ref{30}), corresponding to
the lowest possible order of entanglement between the individual atoms and the
cavity field, may still yield a highly entangled state of the atomic ensemble
after being projected onto a linear combination of the field vacuum and a
highly excited Fock state. Let us take the linear-combination field state%

\begin{equation}
\left\vert F\right\rangle =B\left\vert 0\right\rangle _{p}+C\left\vert
N\right\rangle _{p}, \label{250}%
\end{equation}
where $N$ is the total number of atoms. Then this projection can be shown to
result in the GHZ state. In fact, by casting Eq.(\ref{30}) in the Taylor
series
\begin{equation}
\left\vert \Psi\right\rangle =\frac{1}{A}\overset{N}{\underset{n=1}{%
%TCIMACRO{\dsum }%
%BeginExpansion
{\displaystyle\sum}
%EndExpansion
}}\frac{\left(  \widehat{a}\dagger\right)  ^{n}}{n!}\left(  \sum
\limits_{n=1}^{N}\widehat{\sigma}_{n}^{+}I_{n}\right)  ^{n}\left\vert
\mathcal{O}\right\rangle \label{28}%
\end{equation}
one immediately finds the projection%
\begin{subequations}
\begin{align}
\left\langle F\right.  \left\vert \Psi\right\rangle  &  =\frac{1}{A}\left[
B^{\ast}+\frac{C^{\ast}}{\sqrt{N!}}\left(  \sum\limits_{n=1}^{N}%
\widehat{\sigma}_{n}^{+}I_{n}\right)  ^{N}\right]  \left\vert \emph{O}%
\right\rangle \nonumber\\
&  =\left(  \frac{B^{\ast}}{A}+\frac{C^{\ast}}{A\sqrt{N!}}\prod\limits_{n=1}%
^{N}\widehat{\sigma}_{n}^{+}I_{n}\right)  \left\vert \emph{O}\right\rangle ,
\label{34}%
\end{align}
For
\begin{equation}
B^{\ast}\sqrt{N!}=C^{\ast}\prod\limits_{n=1}^{N}I_{n}, \label{35}%
\end{equation}
this indeed yields the GHZ state with the tanglemeter $f_{GHZ}$ of
Eq.(\ref{tang}). \ 

Detection of the field state $\left\vert F\right\rangle $ cannot however be
directly performed as a probabilistic projection on the photon number basis
and requires a manipulation of the field prior to such a measurement. One way
to perform such a manipulation is to expose the system to the Kerr-nonlinear
interaction Eq.(\ref{27}) with the parameters chosen such that only the
corresponding states with $0$ and $N$ cavity photons are resonant and coupled.
This condition reads
\end{subequations}
\begin{equation}
\left(  \omega_{c}-\omega_{L}\right)  N+\kappa N^{2}=0,
\end{equation}
while the corresponding matrix element of the multiphoton transition between
the photon vacuum and $N$-th Fock state has the form%
\begin{equation}
V_{0N}=\frac{\left(  \kappa\mathcal{E}^{3}\right)  ^{N}\sqrt{N!}}{\overset
{N}{\underset{n=1}{%
%TCIMACRO{\tprod }%
%BeginExpansion
{\textstyle\prod}
%EndExpansion
}}\left(  \kappa n^{2}-\omega_{L}n\right)  }=\frac{\mathcal{E}^{3N}}{\sqrt
{N!}\overset{N}{\underset{n=1}{%
%TCIMACRO{\tprod }%
%BeginExpansion
{\textstyle\prod}
%EndExpansion
}}\left(  n-\omega_{L}/\kappa\right)  }.
\end{equation}
The interaction Eq.(\ref{27}) must be on during a time $t_{Kerr}$, given by
the condition
\begin{equation}
\tan\left(  t_{Kerr}V_{0N}\right)  =\frac{C^{\ast}}{B^{\ast}}=\sqrt{N!}%
/\prod\limits_{n=1}^{N}I_{n},
\end{equation}
Then, if the cavity field is detected in the vacuum state, the projection onto
the state $\left\vert F\right\rangle $ is performed. This results in the GHZ
multiatomic state.

\subsection{Entangling two ensembles}

We now consider entanglement between\ two ensembles of atoms, $A$ and $B$,
each consisting of $N$ identical atoms. Each ensemble is treated as a single
element and we are exclusively concerned with the collective entanglement
between these multiatomic elements expressed in terms of the collective
operators $\widehat{\sigma}_{A}^{+}=\sum\limits_{n=1}^{N}\widehat{\sigma}%
_{n}^{\left(  A\right)  +}$ and $\widehat{\sigma}_{B}^{+}=\sum\limits_{n=1}%
^{N}\widehat{\sigma}_{n}^{\left(  B\right)  +}.$ We guide the first ensemble
of atoms through the cavity and then repeat the procedure for the second
ensemble. If the excitation per atom remains small and the cavity is tuned on
resonance with the atoms frequency,\ then according to Eq.(\ref{EQ18a}) the
combined state of the two multiatomic ensembles and the cavity field reads\
\begin{equation}
\left\vert \Psi\right\rangle =\frac{1}{A}\mathrm{e}^{\mu\widehat{a}%
\dagger\left(  \widehat{\sigma}_{A}^{+}+\widehat{\sigma}_{B}^{+}\right)
}\left\vert \mathcal{O}\right\rangle \ . \label{25}%
\end{equation}
Entanglement between the two ensembles is implied by the presence of cross
terms $\left(  \widehat{\sigma}_{A}^{+}\right)  ^{k}\left(  \widehat{\sigma
}_{B}^{+}\right)  ^{l}$ in the nilpotential, can appear as a result of the
field manipulation prior to the measurement of the cavity photon number. In
accordance with the results of Sect.\ref{methods}, the lowest-order cross term
with $k=l=1$ can be generated\ from the state Eq.(\ref{25}), by squeezing the
cavity field followed by the detection of the cavity vacuum. More
specifically, after the squeezing operator $S=\exp\left(  g\widehat{\alpha
}^{2}-g\widehat{\alpha}^{\dagger2}\right)  $ has been applied to the state
Eq.(\ref{25}) for time $t$, and zero photons are detected, the ensemble state
adopts the form
\begin{subequations}
\begin{equation}
\left\vert \Phi\right\rangle =\frac{r}{A^{\prime}}\exp(\zeta\widehat{O}%
^{2}+\eta)\left\vert \emph{O}\right\rangle \ , \label{32}%
\end{equation}
where%
\begin{align}
\widehat{O}  &  =\mu\left(  \widehat{\sigma}_{A}^{+}+\widehat{\sigma}_{B}%
^{+}\right)  \ ,\nonumber\\
r  &  =\frac{\sqrt{2\pi}}{\sqrt{1+e^{2gt}}},\;\eta=gt,\;\zeta=2\tanh[gt]-gt\;,
\label{38}%
\end{align}
and the value of the parameter $\mu$, given by Eq.(\ref{18a}), is identical
for all the atoms.

\section{Discussion}
\label{V}

We have presented a powerful formalism for the analytical description of
multiatomic ensembles interacting with quantized cavity fields for the case
where the atoms have low excitation probability and the effect of decoherence
is ignored. By combining the analytical functional integration with the
multipartite entanglement description via nilpotent polynomials we have been
able to propose general schemes for the \emph{probabilistic} control of
entanglement among multiple neutral atoms. These techniques consist of
manipulations on the cavity field followed by a projective measurement on the
photon number state and therefore are experimentally feasible only in optical
cavities. We presented several applications in the regime where the frequency
of the cavity mode is resonant with the atomic transition frequency.

The proposed control scheme presumes the resolution of various experimental  
problems posed by state preparation and detection, most importantly the determination of atom number $N$ in the cavity.  There is continuing progress towards the measurement   of  trapped atom numbers, suggesting that this goal may be achieved before long.
The following procedure may  be conceived   of: (\textit{i}) slowly trapping ground-state atoms in the cavity (for which several options exist, including   the adiabatic conveyor - belt technique \cite{Merschede}); (\textit{ii}) counting the trapped atoms by their resonance fluorescence or
other optical techniques; (\textit{iii}) impressing Stark or Zeeman shifts to make atoms at different positions spectrally distinguishable and thus addressable by the
control Raman coupling; (\textit{iv}) switching (on and off)  the Raman control field
in Hamiltonian (\ref{EQ1})  for selected  (spectrally addressable ) atoms for the required interaction time. This procedure, although challenging, may in the long  run allow field - atom entangling manipulations without actually sending atoms in and out of the cavity, but rather making them disappear and reappear by purely optical manipulations.

Beyond the ability to realize a broader variety of entangled multiatom states
than existing states, the main merit of the present analysis is that it has
been exhibited within the unified framework of the  general nilpotent formalism.
\end{subequations}
\section{Acknowledgements}

This work is supported in part by the EC RTN QUACS. A.M. gratefully
acknowledges Ile de France for financial support. G.K. acknowledges the
support of the EC NOE SCALA, ISF and GIF.

\appendix

\section{Functional integration}

The straightforward way to calculate the functional integral of Eq.(\ref{EQ8})
is by transforming it into a Gaussian functional integral. This requires first
to find the stationary solutions $z_{s}^{\ast}(\tau)$, $z_{s}(\tau)$ that
correspond to the extremum of the functional exponent, i.e., the action $S$,%

\begin{align}
S  &  =-\mathrm{i}\int\limits_{0}^{t}\left\vert z(\tau)\right\vert
^{2}\mathrm{d}\tau+\mathrm{i}\sum\limits_{n=1}^{N}\frac{\omega_{n}}%
{2}t\nonumber\\
&  -\mathrm{i}\sum\limits_{n=1}^{N}\widehat{\sigma}_{n}^{+}\int\limits_{0}%
^{t}\mathrm{e}^{\mathrm{i}\omega_{n}\left(  \tau-t\right)  }C_{n}\left(
\tau\right)  z(\tau)\mathrm{d}\tau\nonumber\\
&  +\mathrm{i}\widehat{a}\dagger\int\limits_{0}^{t}\mathrm{e}^{\mathrm{i}%
\omega_{0}(\tau-t)}\mathcal{E}(\tau)z^{\ast}(\tau)\mathrm{d}\tau\nonumber\\
&  -\int\limits_{0}^{t}\int\limits_{\tau}^{t}\mathrm{e}^{\mathrm{i}\omega
_{0}(\tau-\theta)}\mathcal{E}(\tau)\mathcal{E}(\theta)z^{\ast}(\tau)z^{\ast
}(\theta)\mathrm{d}\tau\mathrm{d}\theta, \label{EQ9}%
\end{align}
and then to evaluate the integral for new functional variables $\left\{
z^{\ast}\left(  \tau\right)  +z_{s}^{\ast}(\tau),z\left(  \tau\right)
+z_{s}(\tau)\right\}  $.

\
%TCIMACRO{\TeXButton{TeX field}{\begin{widetext}}}%
%BeginExpansion
\begin{widetext}%
%EndExpansion

By variation of the action $S$
\begin{align}
0  &  =\delta S=\int\limits_{0}^{t}\mathrm{d}\tau\left[  -\mathrm{i}z^{\ast
}(\tau)-\mathrm{i}\sum\limits_{n=1}^{N}\widehat{\sigma}_{n}^{+}C_{n}\left(
\tau\right)  \mathrm{e}^{-\mathrm{i}\omega_{n}\left(  t-\tau\right)  }\right]
\delta z(\tau)\nonumber\\
&  -\left[  \mathrm{i}z(\tau)-\mathrm{i}\widehat{a}\dagger\mathrm{e}%
^{\mathrm{i}\omega_{0}(\tau-t)}\mathcal{E}(\tau)+\mathcal{E}(\tau
)\int\limits_{\tau}^{t}\mathrm{e}^{\mathrm{i}\omega_{0}(\tau-\theta
)}\mathcal{E}(\theta)z^{\ast}(\theta)\mathrm{d}\theta+\mathcal{E}(\tau
)\int\limits_{0}^{\tau}\mathcal{E}(\theta)z^{\ast}(\theta)\mathrm{e}%
^{\mathrm{i}\omega_{0}(\theta-\tau)}\mathrm{d}\theta\right]  \delta z^{\ast
}(\tau)
\end{align}
one obtains the stationary solutions
\begin{align}
z_{s}^{\ast}(\tau)  &  =\mathrm{-}\sum\limits_{n=1}^{N}\widehat{\sigma}%
_{n}^{+}C_{n}\left(  \tau\right)  \mathrm{e}^{-\mathrm{i}\omega_{n}\left(
t-\tau\right)  }\nonumber\\
z_{s}(\tau)  &  =\widehat{a}\dagger\mathrm{e}^{\mathrm{i}\omega_{0}(\tau
-t)}\mathcal{E}(\tau)\nonumber\\
&  -\mathrm{i}\sum\limits_{n=1}^{N}\widehat{\sigma}_{n}^{+}\mathcal{E}%
(\tau)\left(  \int\limits_{\tau}^{t}\mathcal{E}(\theta)C_{n}\left(
\theta\right)  \mathrm{e}^{-\mathrm{i}\omega_{n}\left(  t-\theta\right)
+\mathrm{i}\omega_{0}(\tau-\theta)}\mathrm{d}\theta+\int\limits_{0}^{\tau
}\mathcal{E}(\theta)C_{n}\left(  \theta\right)  \mathrm{e}^{-\mathrm{i}%
\omega_{n}\left(  t-\theta\right)  +\mathrm{i}\omega_{0}(\theta-\tau
)}\mathrm{d}\theta\right)  .\smallskip\label{EQ9a}%
\end{align}%
%TCIMACRO{\TeXButton{TeX field}{\end{widetext}}}%
%BeginExpansion
\end{widetext}%
%EndExpansion
\ 

The substitution of the new variables into Eq.(\ref{EQ9}), separates the
action $S$ into two parts%

\begin{equation}
S=S(z\left(  \tau\right)  ,z^{\ast}\left(  \tau\right)  )+S(z_{s}\left(
\tau\right)  ,z_{s}^{\ast}\left(  \tau\right)  ),
\end{equation}
a \textquotedblleft quantum\textquotedblright\ \ part and a \textquotedblleft
classical\textquotedblright\ one. The quantum contribution of the action does
not contain quantum operators in our case and the functional integration over
this gives just a phase that can be ignored. What is left then is to evaluate
the classical contribution to the functional integral by substituting
\ $S_{cl}=S(z_{s}\left(  \tau\right)  ,z_{s}^{\ast}\left(  \tau\right)  )$
%TCIMACRO{\TeXButton{TeX field}{\begin{widetext}}}%
%BeginExpansion
\begin{widetext}%
%EndExpansion%
\begin{align}
S_{cl}  &  =\mathrm{i}\sum\limits_{n=1}^{N}\frac{\omega_{n}}{2}t-\mathrm{i}%
\widehat{a}\dagger\sum\limits_{n=1}^{N}\widehat{\sigma}_{n}^{+}\int
\limits_{0}^{t}\mathrm{e}^{\mathrm{i}\left(  \omega_{0}+\omega_{n}\right)
(\tau-t)}C_{n}\left(  \tau\right)  \mathcal{E}(\tau)\mathrm{d}\tau\nonumber\\
&  +\sum\limits_{n,m=1}^{N}\widehat{\sigma}_{n}^{+}\widehat{\sigma}_{m}%
^{+}\int\limits_{0}^{t}\int\limits_{0}^{\tau}\mathcal{E}(\tau)\mathcal{E}%
(\theta)C_{n}\left(  \theta\right)  C_{m}\left(  \tau\right)  \mathrm{e}%
^{-\mathrm{i}\left(  \tau(\omega_{0}-\omega_{m}\right)  +t(\omega_{n}%
+\omega_{m)}-\theta(\omega_{n}+\omega_{0}))}\mathrm{d}\theta\mathrm{d}\tau.
\end{align}
into Eq.(\ref{EQ8}). The final expression for the wave function describing the
time-evolution of the combined cavity-atoms state is%
\begin{align}
\left\vert \Psi\left(  t\right)  \right\rangle  &  =\frac{1}{A(t)}\exp\left[
\mathrm{i}\sum\limits_{n=1}^{N}\frac{\omega_{n}}{2}t+\mathrm{i}\widehat
{a}\dagger\sum\limits_{n=1}^{N}\widehat{\sigma}_{n}^{+}\int\limits_{0}%
^{t}\mathrm{e}^{\mathrm{i}\left(  \omega_{0}+\omega_{n}\right)  (\tau-t)}%
C_{n}\left(  \tau\right)  \mathcal{E}(\tau)\mathrm{d}\tau\right] \nonumber\\
&  \exp\left[  +\sum\limits_{n,m=1}^{N}\widehat{\sigma}_{n}^{+}\widehat
{\sigma}_{m}^{+}\int\limits_{0}^{t}\int\limits_{0}^{\tau}\mathcal{E}%
(\tau)\mathcal{E}(\theta)C_{n}\left(  \theta\right)  C_{m}\left(  \tau\right)
\mathrm{e}^{-\mathrm{i}\left(  \tau(\omega_{0}-\omega_{m}\right)
+t(\omega_{n}+\omega_{m)}-\theta(\omega_{n}+\omega_{0}))}\mathrm{d}%
\theta\mathrm{d}\tau\right]  \left\vert \mathcal{O}\right\rangle .
\label{EQ17}%
\end{align}%
%TCIMACRO{\TeXButton{TeX field}{\end{widetext}}}%
%BeginExpansion
\end{widetext}%
%EndExpansion
where $A(t)$ is the normalization factor whose\ derivation is presented in the following.\ 

\subsection{Normalization factor}%

%TCIMACRO{\TeXButton{TeX field}{\begin{widetext}} }%
%BeginExpansion
\begin{widetext}
%EndExpansion
By definition the square of the normalization factor, using the notation of
Eqs.(\ref{EQ18a})-(\ref{18b}), is%
\begin{equation}
\left\vert A(t)\right\vert ^{2}=\left\langle \mathcal{O}\right\vert
\exp\left[  \widehat{a}\sum\limits_{n=1}^{N}\widehat{\sigma}_{n}^{-}%
I_{n}^{\ast}+\sum\limits_{n,m=1}^{N}\widehat{\sigma}_{n}^{-}\widehat{\sigma
}_{m}^{-}I_{m,n}^{\ast}\right]  \exp\left[  \widehat{a}\dagger\sum
\limits_{n=1}^{N}\widehat{\sigma}_{n}^{+}I_{n}+\sum\limits_{n,m=1}^{N}%
\widehat{\sigma}_{n}^{+}\widehat{\sigma}_{m}^{+}I_{n,m}\right]  \left\vert
\mathcal{O}\right\rangle ,
\end{equation}
or, equivalently, after expanding the terms involving the operators
$\widehat{a}$ and $\widehat{a}\dagger$,%
\begin{equation}
\left\vert A(t)\right\vert ^{2}=\left\langle \emph{O}\right\vert \exp\left[
\sum\limits_{n,m=1}^{N}\widehat{\sigma}_{n}^{-}\widehat{\sigma}_{m}^{-}%
I_{m,n}^{\ast}\right]  \exp\left[  \sum\limits_{n=1}^{N}\widehat{\sigma}%
_{n}^{-}I_{n}^{\ast}\sum\limits_{m=1}^{N}\widehat{\sigma}_{m}^{+}I_{m}\right]
\exp\left[  \sum\limits_{n,m=1}^{N}\widehat{\sigma}_{n}^{+}\widehat{\sigma
}_{m}^{+}I_{n,m}\right]  \left\vert \emph{O}\right\rangle . \label{EQ78}%
\end{equation}
Under the assumption of low excitation probability per atom, the collective
operators approximately commute and the Eq.(\ref{EQ78}) may be rewritten in
the form
\begin{equation}
\left\vert A(t)\right\vert ^{2}\simeq\left\langle \emph{O}\right\vert
\exp\left[  \sum\limits_{n,m=1}^{N}\widehat{\sigma}_{n}^{-}\widehat{\sigma
}_{m}^{-}I_{m,n}^{\ast}+\sum\limits_{n=1}^{N}\widehat{\sigma}^{-}I_{n}^{\ast
}\sum\limits_{m=1}^{N}\widehat{\sigma}_{m}^{+}I_{m}+\sum\limits_{n,m=1}%
^{N}\widehat{\sigma}_{n}^{+}\widehat{\sigma}_{m}^{+}I_{n,m}\right]  \left\vert
\emph{O}\right\rangle . \label{EQ79}%
\end{equation}
The expression \ (\ref{EQ79}) can be written in a simpler way as%
\begin{equation}
\left\vert A(t)\right\vert ^{2}=\left\langle \emph{O}\right\vert \exp\left[
\overline{y}My\right]  \left\vert \emph{O}\right\rangle , \label{EQA1}%
\end{equation}
by introducing a Hermitian matrix $M$ and \ a complex vector $y$%
\begin{equation}
M=\left(
\begin{tabular}
[c]{cccccc}%
$\frac{1}{2}I_{1}I_{1}^{\ast}$ & \ldots & $\frac{1}{2}I_{1}I_{N}^{\ast}$ &
$I_{1,1}$ & \ldots & $I_{1,N}$\\
$\vdots$ & $\ddots$ & $\vdots$ & $\vdots$ & $\ddots$ & $\vdots$\\
$\frac{1}{2}I_{N}I_{1}^{\ast}$ & \ldots & $\frac{1}{2}I_{N}I_{N}^{\ast}$ &
$I_{N,1}$ & \ldots & $I_{1,N}$\\
$I_{1,1}^{\ast}$ & \ldots & $I_{N,1}^{\ast}$ & $\frac{1}{2}I_{1}I_{1}^{\ast}$
& \ldots & $\frac{1}{2}I_{N}I_{1}^{\ast}$\\
$\vdots$ & $\ddots$ & $\vdots$ & $\vdots$ & $\ddots$ & $\vdots$\\
$I_{1,N}^{\ast}$ & \ldots & $I_{1,N}^{\ast}$ & $\frac{1}{2}I_{1}I_{N}^{\ast}$
& \ldots & $\frac{1}{2}I_{N}I_{N}^{\ast}$%
\end{tabular}
\ \ \ \ \right)  ,y=\left(
\begin{tabular}
[c]{c}%
$\sigma_{1}^{-}$\\
$\vdots$\\
$\sigma_{N}^{-}$\\
$\sigma_{1}^{+}$\\
$\vdots$\\
$\sigma_{N}^{+}$%
\end{tabular}
\ \ \ \ \right)  .
\end{equation}%
%TCIMACRO{\TeXButton{TeX field}{\end{widetext}}}%
%BeginExpansion
\end{widetext}%
%EndExpansion

At this point, we can employ the identity
\begin{equation}%
%TCIMACRO{\dint }%
%BeginExpansion
{\displaystyle\int}
%EndExpansion
dx^{N}d\overline{x}^{N}\exp\left[  -\left(  \overline{x+w}\right)
A(x+w)\right]  =\frac{\left(  2\pi\right)  ^{N}}{\det A}%
\end{equation}
that holds for complex vectors $x$ and $w$, to prove that%
\begin{equation}
e^{\overline{z}A^{-1}z}=\frac{\det A}{\left(  2\pi\right)  ^{N}}%
%TCIMACRO{\dint }%
%BeginExpansion
{\displaystyle\int}
%EndExpansion
dx^{N}d\overline{x}^{N}\exp\left[  -\overline{x}Ax-\overline{z}x-\overline
{x}z\right]  \ , \label{EQA2}%
\end{equation}
for $A$, a Hermitian matrix and $z$, a vector defined as $z=Aw$.%

%TCIMACRO{\TeXButton{TeX field}{\begin{widetext}}}%
%BeginExpansion
\begin{widetext}%
%EndExpansion

With the help of the identity (\ref{EQA2}) we can rewrite Eq.(\ref{EQA1}) as%
\begin{align}
\left\vert A(t)\right\vert ^{2}  &  =\left\langle \emph{O}\right\vert
\exp\left[  \overline{y}My\right]  \left\vert \emph{O}\right\rangle
\nonumber\\
&  =\frac{\det M^{-1}}{\left(  2\pi\right)  ^{2N}}%
%TCIMACRO{\dint }%
%BeginExpansion
{\displaystyle\int}
%EndExpansion
dx^{2N}d\overline{x}^{2N}\left\langle \emph{O}\right\vert \exp\left[
-\overline{x}M^{-1}x-\overline{y}x-\overline{x}y\right]  \left\vert
\emph{O}\right\rangle \nonumber\\
&  \simeq\frac{\det M^{-1}}{\left(  2\pi\right)  ^{2N}}%
%TCIMACRO{\dint }%
%BeginExpansion
{\displaystyle\int}
%EndExpansion
dx^{2N}d\overline{x}^{2N}\exp\left[  -\overline{x}M^{-1}x+\left(
x+\overline{x}\right)  V\left(  x+\overline{x}\right)  \right] \nonumber\\
&  =\frac{\det M^{-1}}{\left(  2\pi\right)  ^{2N}}%
%TCIMACRO{\dint }%
%BeginExpansion
{\displaystyle\int}
%EndExpansion
dx^{2N}d\overline{x}^{2N}\exp\left[  -\left(
\begin{array}
[c]{cc}%
x & \overline{x}%
\end{array}
\right)  B\left(
\begin{array}
[c]{c}%
\overline{x}\\
x
\end{array}
\right)  \right] \nonumber\\
&  =\frac{\det M^{-1}}{\det B}\ ,
\end{align}
\ where we have assumed that $x_{i}$ small, an assumption valid in the chosen
regime, and%
\begin{equation}
V=\left(
\begin{array}
[c]{cc}%
0 & I_{2N}\\
I_{2N} & 0
\end{array}
\right)  ,\ \ B=\left(
\begin{array}
[c]{cc}%
V & V\\
V & V-M^{-1}%
\end{array}
\right)  \;.
\end{equation}%
%TCIMACRO{\TeXButton{TeX field}{\end{widetext}}}%
%BeginExpansion
\end{widetext}%
%EndExpansion

\section{Squeezing operators \label{appB}}

Having projected the squeezed state onto the vacuum%
\begin{equation}
\left\vert \Phi\right\rangle =\left\langle 0_{p}\right\vert \mathrm{e}%
^{\left(  g\left(  \widehat{\alpha}+\widehat{O}\right)  ^{2}-g^{\ast}%
\widehat{\alpha}^{\dagger2}\right)  t}\left\vert 0_{p}\right\rangle
\mathrm{e}^{\widehat{G}}\left\vert \emph{O}\right\rangle
\end{equation}
we would like to explicitly calculate the state of the atoms in the special
case where $g=g^{\ast}$. \ If the position-momentum representation of the
ladder operations is employed,
\begin{align}
\widehat{\alpha}  &  =\frac{1}{\sqrt{2}}(\widehat{x}+\frac{\partial}%
{\partial\widehat{x}}),\nonumber\\
\widehat{\alpha}^{\dagger}  &  =\frac{1}{\sqrt{2}}(\widehat{x}-\frac{\partial
}{\partial\widehat{x}})
\end{align}
the problem is reduced in solving the Shr\"{o}dinger equation for the
Hamiltonian
\begin{align}
\widehat{H}  &  =i\left(  g\left(  \widehat{\alpha}+\widehat{O}\right)
^{2}-g\widehat{\alpha}^{\dagger2}\right) \nonumber\\
&  =i\left(  \sqrt{2}g\widehat{O}+g\widehat{x}\right)  \frac{\partial
}{\partial\widehat{x}}+i\left(  \sqrt{2}g\widehat{O}\widehat{x}+g+g\widehat
{O}^{2}\right)
\end{align}
and with initial condition
\begin{equation}
\left\langle x|0_{p}\right\rangle =e^{-x^{2}/2}.
\end{equation}
\ If $g$ is a real number, the differential equation to be solved is of first
order both in position and time and thus can be easily handled. The final
result is%
\begin{equation}
\left\vert \Phi\right\rangle =r\exp\left[  \zeta\widehat{O}^{2}+\eta\right]
\exp\left[  \widehat{G}\right]  \left\vert \emph{O}\right\rangle
\end{equation}
where $t$ is time period that the squeezing operator is applied and the
parameters are
\begin{equation}
r=\sqrt{2\pi}/\sqrt{1+e^{2gt}},~~\eta=gt,~~\zeta=2\tanh[gt]-gt.
\end{equation}

\end{document}